\documentstyle[aps,prb,eqsecnum
]{revtex}
\newcommand{\beq}{\begin{equation}}
\newcommand{\eeq}{\end{equation}}
\newcommand{\nn}{\nonumber}

\newcommand{\ie}{{\it i.e.}\ }

\newcommand{\eg}{{\it e.g.}\ }

\begin{document}


\title{Generation of 3rd and 5th harmonics in a thin superconducting film by
       temperature oscillations and isothermal nonlinear\\ current response}

\author{Todor Mishonov$^{a,b}$, Nicolas Ch\'eenne$^{a}$, Didier Robbes$^{c}$ and
Joseph Indekeu$^{a}$\thanks{$\mbox{Corresponding author:}$
$\mbox{phone: (+32 16) 327 127,}$~$\mbox{fax.: (+32 16) 327
983,}$~$\mbox{e-mail:}$~joseph.indekeu@fys.kuleuven.ac.be}}

\address{%
 $^{a}$Laboratorium voor Vaste-Stoffysica en Magnetisme,
    Katholieke Universiteit Leuven,\\
 Celestijnenlaan 200 D, B-3001 Leuven, Belgium\\
 $^b$Department of Theoretical Physics, Faculty of Physics, Sofia University St. Kliment Ohridski,\\ 5 J. Bourchier
    Blvd., 1164 Sofia, Bulgaria\\
 $^{c}$GREYC-Instrumentation, ISMRA,\\6, Bd Mar\'echal Juin, 14050 Caen, France
}
 \preprint{Submitted to Journal of Applied Physics; version of \today}


\maketitle

\begin{abstract}
The generation of harmonics of the voltage response is considered when an AC current is applied through a
superconducting film above $T_c$. It is shown that almost at all temperatures the mechanism of the temperature
oscillations created by the AC current and the temperature dependence of the resistance dominates over the
isothermal nonlinear electric conductivity. Only in a narrow critical region close to $T_c$ the latter is
essential for the generation of the harmonics. A detailed investigation of harmonics generation provides an
accurate method for measuring the thermal boundary conductance between the film and the insulating substrate.
The critical behaviour of the third harmonic will give a new method for the determination of the lifetime of
metastable Cooper pairs above $T_c$. The comparison of the calculated fifth harmonics of the voltage with the
experiment is proposed as an important test for the applicability of the employed theoretical models.
\end{abstract}


\section{Introduction}
\label{sec:introduction}

Third harmonic generation is frequently used in a lot of scientific fields to obtain information which is
impossible to extract from the total signal or even from the first harmonic. Third harmonic techniques are
particularly developed in optics\cite{Barad}, and in the physics of semiconductors\cite{Vandamme,Sauvage}. It is
also possible to explore superconductor properties with such techniques as was already done for thermal properties
in bulk and thin film materials\cite{Robbes,Cheenne,Borca-Tasciuc}, for magnetic response\cite{Raphael},
non-linear effects in YBa$_2$Cu$_3$O$_{7-\delta}$ bicrystal Josephson junctions\cite{Shimakage} and the nonlinear
response in the vortex state\cite{Mallozi}.

The aim of the present paper is to give a theory for the harmonics generation of the voltage response when an AC
current is applied through a conducting film deposited on an insulating substrate. Systematizing the results for
the temperature oscillation mechanism\cite{Robbes,Cheenne} we describe how the effects of isothermal nonlinear
current response can be extracted. Experimental data processing of the harmonic response can give not only an
accurate method for the determination of the thermal boundary conductance $G(T)$ but also the lifetime constant
$\tau_0$ of metastable Cooper pairs above $T_c$. We suppose that the geometry consists of a strip of a superconducting
film.

Generation of 3rd and 5th harmonics in the voltage response of such a sample due to thermal effects of the
electric current is presented in Sec.~\ref{sec:model}. In Sec.~\ref{sec:nonlinear} is performed the comparison of
the thermal mechanism with the isothermal mechanism of nonlinear electric conductivity\cite{MPI} and it is
described how the effects of these two mechanisms can be separated. Finally in Sec.~\ref{sec:discussion} it is
considered how the investigation of the 3rd and 5th harmonics can be used for practical applications in the
physics of bolometers and for the determination of the parameters of superconductors important for understanding
the fundamental processes in these materials.

\section{Model}
\label{sec:model}
%
We consider the temperature $T$ of a strip patterned from a thin superconducting film with length $L$ and width $w$,
deposited on an insulating substrate when a bias current $I$ is applied. Due to ohmic heating, the film
temperature T is slightly higher than the substrate temperature $T_{\rm sub}$:
\beq
\label{Tfilm}
T(t)=T_{\rm sub}+ \Delta T(t).
\eeq
We suppose that the current is low enough to ensure very small temperature variations $\Delta~T\ll~T$. In our
model we will use the phenomenology of the thermal boundary resistance $G(T)$ between film and substrate in which
the total power of emitted phonons is defined as:
\beq
\label{surface T conductance}
P=G\Delta T;
\eeq
for a general review of the thermal boundary resistance between solids see for example the Ref.~\cite{Gmelin} The
power $P$ is proportional to the temperature increase $\Delta T$ and to the total thermal conductance between film
and substrate $G=wLg(T)$, which is equal to the thermal conductivity $g(T)$ multiplied by the cross-section $wL$
of heat exchange. We suppose that the temperature in the film is almost constant across the film in all
directions. Taking into account the thickness of the film $d_{\rm film}$ and thermal conductivity $\kappa$ of the
superconductor, we assume that the thermal boundary resistance is much higher than the thermal resistance of the
thin film ($d_{\rm film}/\kappa\ll1/g$). Due to the insulating nature of the substrate the interface heat exchange
is a purely phononic phenomenon and we expect the thermal boundary conductivity $g(T)$ to be a smooth function of
the temperature, and also $G(T)\approx G(T_{\rm sub})$. We assume that the frequency of the AC current is low
enough, $\omega\ll G(T)/C(T)$, where $C(T)$ is the heat capacity of the film. In this case effects related to the
specific heat are negligible and the temperature follows instantaneously the current through the film, so that
$T=T(I)$. In these conditions, the total power is expressed by Ohm's law
\beq \label{Ohmic heating} P=R(T)I^2,\quad R(T)=\frac{\rho(T)L}{wd_{\rm film}},\quad I=wd_{\rm film}j,\eeq
where $\rho$ and j are respectively the resistivity and the current density. The
comparison with Eq.~(\ref{surface T conductance}) for the emitted power gives the
temperature difference
\beq
\label{DeltaT}
\Delta T=\frac{R(T)}{G(T_{\rm sub})}I^2
 =\frac{\rho(T)d_{\rm film}}{g(T_{\rm sub})}j^{2}.
\eeq
Taking into account that the resistance of the film is temperature dependent $R(T)$
and using Eq.~(\ref{DeltaT}) and Eq.~(\ref{Tfilm}), we easily derive the
power expansion of the resistance as a function of the current:
\begin{eqnarray}
\label{R(T)}
R(T)&=&R(T_{\rm sub}+\Delta T)
=R\left(T_{\rm sub}+\frac{R(T)}{G(T_{\rm sub})}I^2\right)
\approx R\left(T_{\rm sub}+\frac{R(T_{\rm sub}
 +\frac{R(T_{\rm sub})}{G(T_{\rm sub})}I^2)}{G(T_{\rm sub})}I^2\right)\nn\\
&\approx&R+\frac{RR'}{G}I^2+\frac{1}{2}\,\frac{R}{G}\left[2(R')^2+RR''\right]I^4+O(I^6),
\end{eqnarray}
where
\beq
R^\prime(T)=\frac{d}{dT}R(T),\qquad R^{\prime\prime}(T)=\frac{d^2}{dT^2}R(T).
\eeq
Eqs.~(\ref{surface T conductance}-\ref{R(T)}) are essential for the physics of thermo-electric oscillations in
HTSC film structures\cite{Zharov} and the diagnostics of HTSC microbolometers\cite{Korotkov}.
Then for the voltage, we have:
\beq U=R(T)I\approx RI+\frac{RR^{\prime}}{G}I^3
+\frac{1}{2}\,\frac{R}{G^2}\left[2\left(R^{\prime}\right)^2+RR^{\prime\prime}\right]I^5+O(I^7).
\eeq
Let us now consider the case of a harmonic current
\beq
I(t)=I_0\cos\omega t,\qquad j_0=I_0/wd_{\rm film}.
\eeq
Using trigonometric relations
\begin{eqnarray}
\cos^3\omega t&=&\frac{1}{4}\cos3\omega t+\frac{3}{4}\cos\omega t\nn\\
\cos^5\omega t&=&\frac{1}{16}\cos5\omega t+\frac{5}{16}\cos3\omega t+\frac{10}{16}\cos\omega t
\end{eqnarray}
and writing the voltage in a Fourier series
\beq U(t)=U_{1\omega}\cos\omega t+U_{3\omega}\cos3\omega t+U_{5\omega}\cos5\omega
t+\dots, \eeq
we obtain for the amplitude of the harmonics:
\begin{eqnarray}
\label{U135}
U_{1\omega}&=&RI_0+\frac{3}{4}\,\frac{RR^{\prime}}{G}I_0^3
+\frac{5}{16}\,\frac{R}{G^2}\left[2\left(R^{\prime}\right)^2+RR^{\prime\prime}\right]I_0^5+O(I_0^7),\nn\\
U_{3\omega}&=&\frac{1}{4}\,\frac{RR^{\prime}}{G}I_0^3+
 \frac{5}{32}\,\frac{R}{G^2}\left[2\left(R^{\prime}\right)^2+RR^{\prime\prime}\right]I_0^5+O(I_0^7),\nn\\
U_{5\omega}&=&\frac{1}{32}\,\frac{R}{G^2}\left[2\left(R^{\prime}\right)^2+RR^{\prime\prime}\right]I_0^5+O(I_0^7).
\end{eqnarray}
Analogous formulae for the $3\omega$ method applied to thermal conductivity measurements were reported for rod- and
filament-like specimens.\cite{Lu}
For the case of linear dependence of the resistance vs temperature well above $T_c$ for some layered cuprates $R(T)=
U_{1\omega}/I_0\approx AT+B$, these expressions up to $I_0^3$-terms were analyzed in Refs.\cite{Robbes,Cheenne}
in the context of the determination of the thermal boundary conductance $G(T)$ of the
YBa$_2$Cu$_3$O$_{7-\delta}$/SrTiO$_3$ interface explored by the third harmonic generation of the voltage across
the strip. In this case cf.\cite{Robbes,Cheenne} $R'=A$, $R''=0$ and for small enough currents $I_0\ll\sqrt{G/A}$
we have the approximations:
\beq
U_{3\omega}=\frac{AT+B}{4G}AI_0^3,\quad
U_{5\omega}=\frac{AT+B}{16G^2}A^2I_0^5,\quad
G=\frac{U_{1\omega}}{4U_{3\omega}}AI_0^2
 =\frac{U_{3\omega}}{4U_{5\omega}}AI_0^2.
\eeq

For experimental data processing of temperature sweep measurements we can  express all measurements as
functions of the time $t$:
\beq
R'(T)=\frac{\dot{R}}{\dot{T}}=\frac{1}{T'(R)},\qquad
R''(T)=\frac{\ddot{R}\dot{T}-\dot{R}\ddot{T}}{(\dot{T})^3}
=-\frac{T''(R)}{\left(T^{\prime}(R)\right)^3},
\eeq
where a dot means time differentiation; this is helpful if we use numerical filtration of the signal. We suppose
that for small enough currents $I_0\rightarrow 0$, the intensity of the harmonics is decreasing fast with the
overtone index $|U_{1\omega}|\gg|U_{3\omega}|\gg|U_{5\omega}|$. In this case at fixed temperature we can perform a
polynomial fit of $U_{n\omega}/RI_0^n$ versus $I_0^2$, and use at most a second-degree polynomial for
$n=1,\;3,\;5$.

If we investigate the temperature dependence, roughly speaking $U_{1\omega}$ determines the resistance $R(T)$,
and, subsequently, $U_{3\omega}$ the thermal conductance $G(T)$. The comparison of the measured $U_{5\omega}$
with the prediction of Eq.~(\ref{U135}) can be considered as an important test for the consistency of the model.
Using simultaneously the recordings of the first, third and fifth harmonics of the voltage we can obtain better
accuracy for the experimental determination of the resistance and the thermal conductance as follows,
\begin{eqnarray}
\label{RandG}
R(T)&\equiv&\frac{U_{1\omega}-3U_{3\omega}+5U_{5\omega}}{I_0}\approx U_{1\omega}/I_0,\quad\,
\rho(T)\equiv\frac{E_{1\omega}-3E_{3\omega}+5E_{5\omega}}{j_0}\approx E_{1\omega}/j_0,\nn\\
G(T)&\equiv&\frac{1}{4}\,\frac{RR'I_0^3}{U_{3\omega}-5U_{5\omega}}\approx RR'I_0^3/4U_{3\omega},\quad
\;g(T)\equiv\frac{1}{4}\,\frac{d_{\rm film}\rho\rho'j_0^3}{E_{3\omega}-5E_{5\omega}}
\approx d_{\rm film}\rho\rho'j_0^3/4E_{3\omega},
\end{eqnarray}
where $E_{n\omega}\equiv U_{n\omega}/L$ are the amplitudes of the electric fields for $n=1$, 3, 5. These high
precision methods could be helpful for extracting the electric field nonlinearity analyzed in the next section.

\section{Nonlinear conductivity: Numerical example}
\label{sec:nonlinear}
%
Even for negligible temperature oscillations the nonlinear current response of the fluctuation conductivity can
generate harmonics. For superconductors the coefficient ${\cal A}$ in the nonlinear term of the current
\beq
\label{cubic}
j=\frac{E}{\rho(T)}-{\cal A}(T)E^3+O(E^5),\qquad E=\rho\left[j+(\rho j)^3 {\cal A}(T)\right]+O(j^5)
\eeq
describes the depairing effect of the electric field. In other words, an electric field accelerates the metastable
Cooper pairs above $T_c$ and this acceleration increases the decay rate and finally the density of fluctuation
Cooper pairs\cite{MPI}. The nonlinear conductivity coefficient ${\cal A}$ has a strong critical singularity as a
function of the reduced temperature
\beq
\epsilon=\frac{T-T_c}{T_c}\ll1.
\eeq
For a thin film, for example\cite{MPI}, we have
\beq
{\cal A}(\epsilon)
=\frac{\pi^2\xi^2(0)\tau_{\rm rel}^3 e^4}{2^{10}\hbar d_{\rm film}(k_BT_c)^2\epsilon^4},
\eeq
where $e$ is the electron charge, $T_c$ the critical temperature, $\xi(0)$ the coherence length and
$\tau_{\rm rel}$
is the ratio of the lifetime constant $\tau_0$ of the fluctuation Cooper pairs to the theoretical value
\begin{eqnarray}
\label{tau0}
\tau_0^{\rm (theor)}=\frac{\pi}{16}\,\frac{\hbar}{k_BT_c}
\end{eqnarray}
derived within microscopic theory in low-coupling and negligible-depairing
approximation.

According to Ohm's law Eq.~(\ref{Ohmic heating}), and Eqs.~(2.9) and (2.11) both considered mechanisms of 3rd
harmonics generation take the form:
\begin{eqnarray}
E_{3\omega}^{(T)}&\equiv&\frac{U_{3\omega}^{(T)}}{L}
\approx \frac{1}{4}\rho j_0^3\;\left(\frac{d_{\rm film}\rho'}{g}\right),\quad\;
E_{3\omega}^{(E)}\equiv\frac{U_{3\omega}^{(E)}}{L}
\approx\frac{1}{4}\rho j_0^3\left(\rho^3{\cal A}\right)\nn\\
U_{3\omega}&=&L\left(E_{3\omega}^{(T)}+E_{3\omega}^{(E)}\right)
 \approx \frac{1}{4}\rho L\left(\frac{I_0}{d_{\rm film}w}\right)^3\,
 \left(\frac{d_{\rm film}\rho'}{g}+\rho^3{\cal A}\right),
\end{eqnarray}
where $\rho'(T)=d\rho(T)/dT$ and the superscript denotes the thermal $(T)$ or electric-field $(E)$ origin of the
harmonics generation. $U_{3\omega}$ is the total voltage signal measured by the lock-in. It consists of two
contributions: the thermal part is proportional to the thickness of the film because heat is generated in the bulk
of the film but leaves it through the interface; the electric part is thickness-independent. The former behaves
near $T_c$ as a constant with a correction proportional to $\epsilon^{-2}$, while the latter displays a strong
singularity proportional to $\epsilon^{-4}$, as given in Eq.~(3.3).

Let us make an order of magnitude evaluation taking the variable $T$ to $T_c$ and taking $\tau_0$ as given by
Eq.~(\ref{tau0}), i.e. $\tau_{\rm rel}=1$:
\beq
\frac{U_{3\omega}^{(E)}}{U_{3\omega}^{(T)}}\approx\left(\frac{\epsilon_{ET}}{\epsilon}\right)^4,\qquad
\epsilon_{ET}\equiv\left(\frac{\pi^2e^4g(T_c)\xi^2(0)\rho_N^3(T_c)}
{2^{10}\hbar\rho'_N(T_c)(k_BT_cd_{\rm film})^2}\right)^{1/4}, \eeq
where $\rho_N(T_c)$ is the extrapolated normal conductivity from temperatures well above $T_c$.
For high-$T_c$ cuprates\cite{Nahum,Dousselin} we have
\begin{eqnarray}
g(T_c)&\approx&1000\;{\rm W/K\,cm^2},\quad \xi(0)=1.1\;{\rm nm},\quad
\rho_N(100\,{\rm K})=132\;{\rm \mu\Omega\, cm},\quad\nn\\
\rho'_N(100\,{\rm K})&=&1.09\;{\rm \mu\Omega\, cm/K},\quad T_c=90\;{\rm K},\quad d_{\rm film}=50\;{\rm nm},\quad
\epsilon_{ET}\simeq1\%.
\end{eqnarray}
This means that the nonlinear effect due to the electric field will dominate in a 1~K region around $T_c$, and
will still be observable 2 or 3~K above $T_c$. In order to extract this term we have to determine $G(T)$ from
Eq.~(\ref{RandG}), using the experimental data in the temperature interval, \eg, $(1.1\,T_c,\;2\,T_c)$, well above
the critical region. Being a purely phononic process for insulating substrates the thermal conductivity $g(T)$ has
no critical singularities and can be reliably extrapolated to $T_c$ using the polynomial fit of the experimental
data. Then this polynomial approximation $G_{\rm fit}(T)=Lwg_{\rm fit}(T)$ can be used in Eq.~(\ref{U135})
together with the data for the resistance in order to compare the experimental data with the prediction from the
model of the thermal oscillations. The model has two functions $g(T)$ and $\rho(T)$ which have to be
experimentally determined and fitted. We suppose that the simple model of purely thermal oscillations will describe
the main qualitative property of the temperature-dependence of the 3rd harmonic -- the maximum at $T_c$
determined by $\rho'.$ This maximum already contains a critical behaviour due to the critical behaviour of the
conductivity $1/\rho(T)$ which has a significant fluctuation part close to $T_c.$ The singularity created by the
depairing effects, however, is stronger close to the critical region and we can determine this electric
nonlinearity using the experimental data and the fitted thermal conductance. Taking into account Eqs.~(\ref{U135}) and
(\ref{cubic}) we derive the experimental definition of the nonlinear voltage and its coefficient
\begin{eqnarray}
U_{3\omega}^{(E)}(T)&\approx& U_{3\omega}(T)
 -RI_{3\omega}^{\rm (gen)}
 -\frac{1}{4}\,\frac{RR^{\prime}}{G_{\rm fit}}I_0^3
 -\frac{5}{32}\,\frac{R}{G_{\rm fit}^2}\left[2\left(R^{\prime}\right)^2+RR^{\prime\prime}\right]I_0^5
 =\frac{1}{4}L\rho^4{\cal A}j_0^3,\nn\\
{\cal A}_{\rm exper}(T)&\approx&\frac{4L^3}{wd_{\rm film}R^4I_0^3}
 \left\{U_{3\omega}-RI_{3\omega}^{\rm (gen)}
 -\frac{RR'I_0^3}{4G_{\rm fit}}-\frac{5RI_0^5}{32G_{\rm fit}^2}
 \left[2\left(R^{\prime}\right)^2+RR^{\prime\prime}\right]\right\},
\end{eqnarray}
where $I_{3\omega}^{\rm (gen)}$ is the small parasite amplitude of the 3rd harmonic of the current created by the
current generator. The current $I_0$ should be small enough
\beq
I_0\ll\sqrt{\frac{G_{\rm fit}}{\left|R'\right|}},
\eeq
in order for the $I_0^5$ term to be much smaller than the $I_0^3$ term, and for the $I_0^7/G_{\rm fit}^3$
correction to be negligible. The nonlinear coefficient determined in this way should be compared with the results of the
fluctuation theory for a layered superconductor\cite{MPI}
\beq
{\cal A}_{\rm theor}(\epsilon)=\frac{4k_BTe^4\left[\xi_a(0)\tau_0\right]^3}{\pi\hbar^4s\xi_b(0)}\,
\frac{\left[\epsilon^3+\frac{3}{2}r\epsilon^2+\frac{9}{8}r^2\epsilon+\frac{5}{16}r^3\right]}
{\left[\epsilon(\epsilon+r)\right]^{7/2}},
\eeq
where $\xi_a(0)$, $\xi_b(0)$  and  $\xi_c(0)$ are the coherence lengths, $s$ is the interlayer distance, and
$r\equiv(2\xi_c(0)/s)^2$. We suppose that the electric field is applied along the $a$-axis. For high-$T_c$
cuprates a significant nonlinear effect can be observed only for very thin high-quality films.

Very strong nonlinear $I-V$ characteristics $V\propto I^3$ are observed\cite{FioryHebardGlaberson,Matsuda} at
Kosterlitz-Thouless transition temperature $T_{\rm KT}$. At this temperature 3rd harmonic will have a
$\lambda$-shaped maximum analogous to the maximum observed at 1st harmonic\cite{Fiory_etal}. For temperatures
slightly above $T_{\rm KT}$ the 3rd harmonic is created mainly by fluctuation Cooper pairs. Except in the
considered narrow critical region the thermal oscillation mechanism will give the main part of the harmonics
generation in the general case. Only for some conventional superconductor films the $\epsilon_{ET}$ parameter may
be large enough so that the thermal contribution can be neglected but this problem requires additional
investigation.

\section{Discussion and conclusions}
\label{sec:discussion}
%
Looking at the expression of $\epsilon_{ET}$, it appears that in high-$T_c$ superconductors the electric field
nonlinearity for the generation of the third harmonic can be with better accuracy measured for high-quality films
as thin as possible. In this case the investigation of the third harmonic between $T_c$ and $T$ a few degrees
above $T_c$ can be used for the determination of the lifetime constant $\tau_0$ of fluctuation Cooper pairs.

Far above the critical region, we considered a quite universal mechanism of the voltage harmonics generation due
to thermal oscillations. This mechanism is very effective close to the phase transition, superconducting or
metal-insulator, where the resistivity changes significantly in a narrow temperature interval. Perovskite thin
films, high-$T_c$ cuprates and manganates with colossal magnetic resistance are very promising for technical
applications like bolometry\cite{Robbes,Cheenne}, for example. Detailed investigation of thermal properties by
simple measurement of 3rd harmonic can give useful information about the quality of the sample. A detailed
investigation of the temperature dependence of harmonic generation can give a precise method for the
determination\cite{Robbes,Cheenne} of the thermal boundary conductivity $g(T)$. Very intense third harmonic
generation could be expected for Bi$_2$Sr$_2$CaCu$_2$O$_8$ whiskers placed in vacuum. The black body radiation is
analogous to the phonon radiation in the insulator but the emitted power is much smaller, because at low
temperatures the power is inversely proportional to the square of the wave velocity in the heat sink.

In such a way the fundamental problem of the physics of superconductivity, the determination of the Cooper pair
lifetime constant and the problem of the determination of the thermal boundary conductivity $g(T)$, important for
many technical applications, can be solved simultaneously by investigating the voltage harmonics generation in
superconducting microbridges.

\acknowledgments

We have highly appreciated interesting discussions with K.~Maki and A.~Varlamov and have benefitted from  their
stimulating comments. This research has been supported by the Belgian DWTC, IUAP, the Flemish GOA and VIS/97/01
Programmes. T.M. is KUL Senior Fellow (F/00/038) and N.Ch. is supported by the EC TMR Network Contract
nr. ERB-FMRX-CT-980171.


\newpage
\section*{Appendix}
\label{sec:appendix}
%
In this appendix we will consider in short some formulae, convenient for programming, for smoothing and numerical
differentiation of a function measured experimentally $f_j=f(t_j)$ for $j=1,\;2,\;3,\;\dots,\;N$ with some error.
The main idea is to perform in the beginning the interpolation of the function using the Taylor expansion at some
point $t$ and the values of the function averaged with respect to the experimental data and its 1st and 2nd
derivatives $\langle f(t)\rangle$, $\langle\dot f(t)\rangle$ and $\langle\ddot f(t)\rangle$
\beq
\label{parabola}
f(\tilde t)\approx\langle f(t)\rangle+\langle\dot f(t)\rangle(\tilde t-t)
 +\frac{1}{2}\langle\ddot f(t)\rangle(\tilde t-t)^2.
\eeq
Applying this parabolic interpolation to the interpolation points $\tilde t=t_i$ we will determine the averaged
values of the function searching the minimum of the sum of squares of the errors
\beq
G(\langle f\rangle, \langle\dot f\rangle, \langle\ddot f\rangle)=
\sum_{k=1}^{N}
\left[
\langle f\rangle+\langle\dot f\rangle(t_k-t)
 +\frac{1}{2}\langle\ddot f\rangle(t_k-t)^2-f_k
\right]^2w_k,
\eeq
where the weights $w_k>0$ depend on the distance from the interpolation point $t_k$ to the center of the
interpolation $t$, and for simplicity we skip this argument in the notation for the averaged function $\langle
f\rangle=\langle f(t)\rangle,\; \dots$. One possible weight function is
\beq
w_k=\frac{1}{\exp[(|t_k-t|-W_\tau)/S_\tau]+1}
\eeq
where $W_\tau$ is the half-width of the interpolation interval and $S_\tau$ is the sharpness of this window. The
minimum of the Gaussian function $G$ with respect of $\langle f\rangle,$ $\langle \dot f\rangle$, $\langle \ddot
f\rangle$ gives the equations
\beq
\left(
\begin{array}{lll}
          \left[1\right]&\left[\tau\right]&\left[\tau^2\right]\\
          \left[\tau\right]&\left[\tau^2\right]&\left[\tau^3\right]\\
          \left[\tau^2\right]&\left[\tau^3\right]&\left[\tau^4\right]\\
\end{array}
\right)
\left(\begin{array}{l}\langle f\rangle\\\langle \dot f\rangle\\\frac{1}{2}\langle \ddot f\rangle\end{array}\right)
=\left(\begin{array}{l}
               \left[ f \right] \\  \left[ \tau f \right] \\ \left[\tau^2 f\right]\\
         \end{array}\right)
\eeq
which have the solution
\beq
\label{averaging}
\langle f\rangle=\frac{D_x}{D},\qquad \langle\dot f\rangle=\frac{D_y}{D},\qquad
\frac{1}{2}\langle\ddot f\rangle=\frac{D_z}{D},
\eeq
where
\begin{eqnarray}
D&=&[1][\tau^2][\tau^4]+2[\tau][\tau^2][\tau^3]-[\tau^2]^3-[\tau]^2[\tau^4]-[1][\tau^3]^2,\nn\\
D_x&=&[f]([\tau^2][\tau^4]-[\tau^3]^2)+[\tau f]([\tau^2][\tau^3]
     -[\tau][\tau^4])+[\tau^2f]([\tau][\tau^3]-[\tau^2]^2)\nn,\\
D_y&=&[f]([\tau^2][\tau^3]-[\tau][\tau^4])+[\tau f]([1][\tau^4]-[\tau^2]^2)
     +[\tau^2 f]([\tau][\tau^2]-[1][\tau^3])\nn,\\
D_z&=& [f]([\tau][\tau^3]-[\tau^2]^2)+[\tau f]([\tau][\tau^2]-[1][\tau^3])+[\tau^2f]([1][\tau^2]-[\tau]^2),
\end{eqnarray}
and
\beq
\tau_k=(t_k-t),\qquad [\tau^nf]=\sum_{k=1}^{N} \tau_k^n f_kw_k,\qquad
[\tau^n]=\sum_{k=1}^N \tau_k^nw_k,\qquad [1]=\sum_{i=1}^Nw_k,\qquad n=0,\,2,\,3,\,4.
\eeq
The formulae simplify significantly if we use equidistant points $t_j=j\Delta t$ and take the center of the
interpolation at one of them $t=t_i$, \ie
\beq
\label{nodes}
f_{i+k}\approx\langle f_i\rangle+\langle\dot f_i\rangle \Delta t k
 +\frac{1}{2}\langle\ddot f_i\rangle(\Delta t)^2k^2.
\eeq
In this case the odd momenta are zero, $[\tau]=0$ and $[\tau^3]=0$, and we have
\begin{eqnarray}
D&=&[1][\tau^2][\tau^4]-[\tau^2]^3,\nn\\
D_x&=&[f][\tau^2][\tau^4]-[\tau^2f][\tau^2]^2,\nn\\
D_y&=&[\tau f]([1][\tau^4]-[\tau^2]^2),\nn\\
D_z&=&[\tau^2f]([1][\tau^2]-[f][\tau^2]^2).
\end{eqnarray}
A further simplification is to take a sharp window for interpolation
\beq
W_\tau=\left(L+\frac{1}{2}\right)\Delta t,\qquad S_\tau=0,\qquad
w_k=\left\{\begin{array}{l}1, \mbox{ for } |k|\le L\\0, \mbox{ for } |k| > L\end{array}\right..
\eeq
This means that sums include only $2L+1$ points centered at $i$ and we can use directly the integer index $k$
instead of the argument $\tau$
\begin{eqnarray}
D&=&[1][k^2][k^4]-[k^2]^3,\nn\\
D_x&=&[f][k^2][k^4]-[k^2f][k^2]^2,\nn\\
D_y&=&[k f]([1][k^4]-[k^2]^2),\nn\\
D_z&=&[k^2f]([1][k^2]-[f][k^2]^2),
\end{eqnarray}
where
\beq
[k^nf]=\sum_{k=-L}^{L} k^n f_{i+k},\qquad
[k^n]=\sum_{k=-L}^{L} k^n,\qquad [1]=\sum_{k=-L}^{L}1=2L+1,\qquad n=0,\,2,\,3,\,4.
\eeq
The simplest possible example is probably the 7-point averaging for $L=3$, in this case for $2<i<N-2$
Eq.~(\ref{averaging}) gives:
\begin{eqnarray}
\langle f_i\rangle&=&\frac{1}{21}\left[-2(f_{i-3}+f_{i+3})+3(f_{i-2}+f_{i+2})+6(f_{i-1}+f_{i+1})+7f_i\right],\nn\\
\langle \dot f_i\rangle&=&\frac{1}{28\Delta t}\left[3(f_{i+3}-f_{i-3})+2(f_{i+2}-f_{i-2})+(f_{i+1}-f_{i-1})\right],\nn\\
\langle \ddot f_i\rangle&=&\frac{1}{42(\Delta t)^2}\left[5(f_{i-3}+f_{i+3})-3(f_{i-2}+f_{i+2})+4f_i\right].
\end{eqnarray}
Some of these formulae can be used sequentially for smoothing, differentiation and further smoothing of the derivative. For
the ends we can apply the parabolic interpolation Eq.~(\ref{nodes})
\beq
\langle f_{k}\rangle=\langle f_i\rangle+\langle\dot f_i\rangle \Delta t (k-i)
 +\frac{1}{2}\langle\ddot f_i\rangle(\Delta t)^2(k-i)^2,
\eeq
for $i=3$ and $k=1,\;2$. The same formula has to be applied for the right end of the data for $i=N-3$ and
$k=N-1,\;N-2$. Finally the parabolic interpolation Eq.~(\ref{parabola}) can be used for tabulation of the
inverse function $t(f)$
\beq
\tilde t\approx t + \frac{\langle\dot f(t)\rangle}{\langle\ddot f(t)\rangle}
\left[\sqrt{\frac{2\langle\ddot f(t)\rangle(f-\langle f(t)\rangle)}{\langle\dot f(t)\rangle^2}+1}-1\right].
\eeq
\end{document}